\documentstyle[12pt]{article}
\input psfig.sty
\def\nn{\noindent}
\def\Re{{\cal R \mskip-4mu \lower.1ex \hbox{\it e}\,}}
\def\Im{{\cal I \mskip-5mu \lower.1ex \hbox{\it m}\,}}
\def\ie{{\it i.e.}}

\def\sub#1{_{\lower.25ex\hbox{$\scriptstyle#1$}}}
\def\tev{\,{\ifmmode\mathrm {TeV}\else TeV\fi}}
\def\gev{\,{\ifmmode\mathrm {GeV}\else GeV\fi}}
\def\mev{\,{\ifmmode\mathrm {MeV}\else MeV\fi}}
\def\mpl{\ifmmode \overline M_{Pl}\else $\overline M_{Pl}$\fi}
\def\to{\rightarrow}

\def\subw{_{\rm w}}
\def\mh{\ifmmode m\sbl H \else $m\sbl H$\fi}
\def\mch{\ifmmode m_{H^\pm} \else $m_{H^\pm}$\fi}
\def\mt{\ifmmode m_t\else $m_t$\fi}
\def\mc{\ifmmode m_c\else $m_c$\fi}
\def\mz{\ifmmode M_Z\else $M_Z$\fi}
\def\mw{\ifmmode M_W\else $M_W$\fi}
\def\mws{\ifmmode M_W^2 \else $M_W^2$\fi}
\def\mhs{\ifmmode m_H^2 \else $m_H^2$\fi}   
\def\mzs{\ifmmode M_Z^2 \else $M_Z^2$\fi}
\def\mts{\ifmmode m_t^2 \else $m_t^2$\fi}
\def\mcs{\ifmmode m_c^2 \else $m_c^2$\fi}
\def\mchs{\ifmmode m_{H^\pm}^2 \else $m_{H^\pm}^2$\fi}
\def\ztwo{\ifmmode Z_2\else $Z_2$\fi}
\def\zone{\ifmmode Z_1\else $Z_1$\fi}
\def\mtwo{\ifmmode M_2\else $M_2$\fi}
\def\mone{\ifmmode M_1\else $M_1$\fi}
\def\tb{\ifmmode \tan\beta \else $\tan\beta$\fi}
\def\xw{\ifmmode x\subw\else $x\subw$\fi}
\def\ch{\ifmmode H^\pm \else $H^\pm$\fi}
\def\lum{\ifmmode {\cal L}\else ${\cal L}$\fi}
\def\inpb{\,{\ifmmode {\mathrm {pb}}^{-1}\else ${\mathrm {pb}}^{-1}$\fi}}
\def\infb{\,{\ifmmode {\mathrm {fb}}^{-1}\else ${\mathrm {fb}}^{-1}$\fi}}
\def\epem{\ifmmode e^+e^-\else $e^+e^-$\fi}
\def\ppb{\ifmmode \bar pp\else $\bar pp$\fi}
\def\bsg{\ifmmode B\to X_s\gamma\else $B\to X_s\gamma$\fi}
\def\bsll{\ifmmode B\to X_s\ell^+\ell^-\else $B\to X_s\ell^+\ell^-$\fi}
\def\bstt{\ifmmode B\to X_s\tau^+\tau^-\else $B\to X_s\tau^+\tau^-$\fi}
\def\lamt{\ifmmode \tilde\lambda\else $\tilde\lambda$\fi}
\def\shat{\ifmmode \hat s\else $\hat s$\fi}
\def\that{\ifmmode \hat t\else $\hat t$\fi}
\def\uhat{\ifmmode \hat u\else $\hat u$\fi}

\newskip\zatskip \zatskip=0pt plus0pt minus0pt
\def\matth{\mathsurround=0pt}

\def\atversim#1#2{\lower0.7ex\vbox{\baselineskip\zatskip\lineskip\zatskip
  \lineskiplimit 0pt\ialign{$\matth#1\hfil##\hfil$\crcr#2\crcr\sim\crcr}}}

\renewcommand{\thefootnote}{\fnsymbol{footnote}}

\begin{document} \begin{titlepage} 
\rightline{\vbox{\halign{&#\hfil\cr
&SLAC-PUB-8636\cr
&FERMILAB-Pub-00/254-T\cr
&October 2000\cr}}}
\begin{center}

{\Large\bf Gravi-Burst: \\
 Super-GZK Cosmic Rays from Localized Gravity}
\footnote{Work supported by the Department of 
Energy, Contract DE-AC03-76SF00515}
\medskip

\normalsize 
{\bf H. Davoudiasl$^a$, J.L. Hewett$^{a,b}$, and T.G. Rizzo$^a$}
\vskip .3cm
$^a$Stanford Linear Accelerator Center \\
Stanford University \\
Stanford CA 94309, USA\\
\vskip .2cm
$^b$Fermi National Accelerator Laboratory \\
Batavia IL 60510, USA\\
\vskip .3cm

\end{center}

\begin{abstract} 
The flux of cosmic rays beyond the GZK cutoff ($\sim 10^{20}$ eV) may be 
explained through their production by ultra high energy cosmic neutrinos, 
annihilating on the relic neutrino background, in the vicinity of 
our galaxy.  This process is mediated through the production of a $Z$ boson 
at resonance, and is generally known as the $Z$-Burst mechanism.  We show 
that a similar mechanism can also contribute to the super-GZK spectrum at 
even higher, ultra-GZK energies, where the particles produced at resonance 
are the Kaluza-Klein gravitons of weak scale mass and coupling from the 
Randall-Sundrum (RS) hierarchy model of localized gravity model. We call this 
mechanism Gravi-Burst.  We discuss the parameter space of relevance to 
Gravi-Bursts, and comment on the possibility of its contribution to the 
present and future super-GZK cosmic ray data and place bounds on the 
RS model parameters. Under certain assumptions about the energy spectrum of 
the primary neutrinos we find that cosmic ray data could be potentially as 
powerful as the LHC in probing the RS model. 
\end{abstract} 




\renewcommand{\thefootnote}{\arabic{footnote}} \end{titlepage}


\section{Introduction}

About 25 years ago, Greisen, Zatsepin, and Kuzmin (GZK) noted that the 
observed spectrum of proton, photon,
and nucleus cosmic rays must virtually end at energies above $\sim 10^{20}$ 
eV, the GZK cutoff \cite{GZK}.  Their
key observation was that Ultra High Energy Cosmic Rays (UHECR's) deplete 
their energy through various
interactions with the 2.7$^{\, ^\circ}$ K Cosmic Microwave Background 
Radiation (CMBR), over distances of order
$10-100$ Mpc.  Above $10^{19}$ eV, nuclei are photo-dissociated by 
interactions with the CMBR, and a $10^{20}$
eV proton loses most of its energy over a distance of $\sim 50$ Mpc.  The 
analogous distance for a photon of the
same energy is $\sim 10$ Mpc, due to $e^+ e^-$ pair production on the 
radio background \cite{radio}.

However, over the past three decades, different experiments have observed a 
total of about 20 events at or above this 
$10^{20}$ eV bound\cite{20sgzk}.  Since there seem to be no feasible candidates 
for the sources of these cosmic rays,
such as Active Galactic Nuclei, within a GZK distance $\sim 50$ Mpc of the 
earth, the observation of these events
poses a dilemma.  A number of proposals have been made to resolve this 
puzzle \cite{Olinto}.  One such
proposal for the origin of the super-GZK events, due to Weiler, is based on 
the observation that UHECR
neutrinos can travel over cosmological distances, with negligible energy 
loss{\cite{Zburst,Weiler}}.  Therefore, if these neutrinos are
present in the universe they could in principle produce $Z$ bosons on 
resonance through annihilation on the relic
neutrino background, within a GZK distance of the earth.  The highly boosted 
subsequent decay products of the
$Z$ will then be observed as primaries at super-GZK energies, since they do 
not have to travel cosmological
distances to reach us.  This mechanism for producing super-GZK cosmic rays 
is referred to as $Z$-Burst.

The $Z$-burst mechanism has the advantage that it does not assume physics 
beyond the Standard Model (SM)
and is, therefore, minimalistic.  However, any extension of the SM that 
provides a particle $X$ which couples to
$\nu {\bar \nu}$ and decays into the usual primaries can in principle 
contribute to the super-GZK spectrum beyond the range presently observed. 
Assuming a mass $m_\nu \sim 10^{-2}-10^{-1}$ eV for neutrinos as suggested by 
atmospheric oscillation data, the particle $X$ 
must have a mass of order the weak
scale ($\sim 1$ TeV) to be relevant to the spectrum near the GZK cutoff.  In 
this paper, we will show that the
massive Kaluza-Klein (KK) tower of gravitons in the Randall-Sundrum (RS) 
localized gravity model{\cite {RS}} are viable candidates for particle $X$.

The RS model is based on a truncated five-dimensional Anti-deSitter($AdS_5$) 
spacetime, with two 4-$d$ Minkowski boundaries.  Our visible
4-$d$ universe and all fields associated with the SM are 
assumed to be confined on one of these boundaries, referred to as the TeV 
brane, with the other `Planck' brane
boundary separated from us by a fixed
distance $r_c \sim 10 \, \mpl^{-1}$, the compactification scale 
along the $5^{th}$ dimension;  
$\mpl$ is the reduced Planck mass.   The RS geometry is such that the 
induced metric on the visible TeV brane 
generates the weak scale from a 5-$d$ scale $M_5 \sim \mpl$, without 
fine-tuning, through an
exponentiation.  The interested reader is referred to Refs.  \cite{RS,dhrprl} 
for the details of the RS model and its
numerous phenomenological implications.  However, here we mention that a 
distinct feature of this model is that
it predicts the existence of a tower of spin-2 KK gravitons, 
$G^{(n)}$ ($n = 1, 2, 3, \ldots$), starting at the weak
scale, and with weak scale mass splittings and couplings.

Phenomenological studies \cite{dhrprl,dhrprd} suggest that the lowest lying KK 
graviton $G^{(1)}$ can be as light as $\sim
400$ GeV.  The $G^{(n)}$ have couplings to all particles, due to their 
gravitational origin and can be
produced by $\nu {\bar \nu}$ annihilation, eventually decaying into 
$q {\bar q}, g g, \gamma \gamma, \ldots$. 
Thus, the $G^{(n)}$ can in principle contribute to the super-GZK spectrum in a 
way that is similar to the $Z$-burst
contribution.  We call this graviton mediated process Gravi-Burst.

Since the $Z$ and $G^{(n)}$ have different couplings and branching fractions 
to the observed primary particles,
we expect that experiments may be able to distinguish between $Z$-burst and 
gravi-burst initiated primaries. 
Also, depending on the behavior of the flux of neutrinos at super-GZK 
energies, more than one member of the KK
graviton tower could contribute to gravi-burst.  In this case, the RS 
model predicts a characteristic
multi-peaked behavior for future data at super-GZK energies and beyond.  
However, collider experiments may be
a better place to directly search for the graviton tower, with cosmic ray 
data providing complementary information as we will discuss in detail later.

In the next section, we present the necessary formulae for estimating the  
super-GZK flux in the $Z$-burst
model.  We adapt this approach to gravi-burst and give the corresponding 
rate estimates in this scenario.  Section 3
contains our results for a range of RS model parameters and a comparison 
with the $Z$-burst predictions.  We will show that if the neutrino spectrum 
falls sufficiently slowly 
with energy we can use GZK data to greatly restrict the 
parameter space of the RS model. 
Our conclusions are given in section 4.

\section{The Burst Mechanism}

The burst mechanism relies on several well-motivated 
assumptions given the successes of the SM, Big Bang Nucleosynthesis 
and the observation of neutrino oscillations due to the existence of finite 
neutrino masses. This scenario is most easily demonstrated in terms of 
the conventional $Z$-burst. 
This model proposes that a high energy flux of neutrinos (and anti-neutrinos) 
are produced by some as yet unknown astrophysical source and 
collide with the relic background neutrinos in the 
galactic neighborhood. The origin of this flux is unspecified but constraints 
on its magnitude and energy dependence exist from Fly's Eye 
data{\cite {fluxcons}}. If the flux at the $Z$-pole is sufficient to explain 
the super-GZK excess then the Fly's Eye data tells us that the fall off with 
energy of the neutrino flux at somewhat lowers energies goes at least as 
fast at $E^{-0.9}$. A similar energy behavior may be expected above the 
$Z$-pole. 

Due to the finiteness of neutrino masses one would expect that the local 
density of neutrinos will most likely be enhanced over the uniform 
cosmological background 
due to their gravitational clustering around the galaxy{\cite {Weiler}}. 
Massive neutrinos 
within a few $Z$ widths of the right energy 
\begin{equation}
E^{R_Z}_\nu=M_Z^2/2m_\nu=4(0.1~{\rm eV}/m_\nu)\times 10^{22}~{\rm eV} \,,
\end{equation}
will then resonantly annihilate into hadrons with the local anti-neutrinos 
(and vice versa) at the $Z$-pole with the large cross section
\begin{equation}
<\sigma_{ann}>^Z=\int {ds\over {M_Z^2}} \sigma_{ann}(s)=
{4\pi G_F B^Z_h\over {\sqrt 2}}\simeq 28{\rm nb} \,,
\end{equation}
where $B^Z_h\simeq 0.70$ is the hadronic branching fraction of the $Z$. We 
assume that only left-handed neutrinos exist and employ the narrow width 
approximation. Given a neutrino mass hierarchy and the Super-Kamiokande 
atmospheric oscillation results{\cite {SuperK}} we 
expect one of the neutrinos to have a mass near $\simeq 0.05-0.06$ eV. (This 
follows from using the latest two parameter fit to the Super-K 
data which yields a 
value for $\Delta m^2$ of $3.2\times 10^{-3}$ eV$^2$ 
and by supposing that one of the 
neutrino masses is at least a few times  larger than the second.) 
The locally produced 30 or so hadrons from the decay of the $Z$ 
are then the effective primaries for the super-GZK events that are observed 
with energies in excess of 
$\sim 10^{20}$ eV. (In principle, there being three neutrinos, we should 
consider three different cases depending on their masses. This is a 
straightforward extension of the present discussion.) 
If the source of the initial neutrinos is randomly 
distributed in space then, as shown by Weiler{\cite {Weiler}}, we can 
calculate the total rate 
of super-GZK events induced by $\nu-\bar \nu$ annihilation at the $Z$ pole 
within a distance $D$ of the Earth as 
\begin{equation}
F_Z\simeq E^{R_Z}_\nu F_\nu(E^{R_Z})<\sigma_{ann}>^Z\int_0^D~dx ~n(x) \,,
\end{equation}
where the narrow width approximation has again been employed, $F_\nu(E^{R_Z})$ 
is the incident neutrino flux evaluated at the resonant energy, and $n(x)$ 
is the column number density of neutrinos. In deriving this expression it is  
assumed  that the product 
$<\sigma_{ann}>^Z\int_0^D~dx ~n(x)<<1$ as is the case for the $Z$ in the SM 
and in the RS model we consider below. In practice we are interested in 
rather close annihilation, \ie, values 
of $D$ of order the GZK limit for protons which is $\sim 50$ Mpc. 

Weiler has shown that for reasonable ranges of the parameters the resulting 
value of the flux $F_Z$ can indeed explain the $\simeq 20$ events 
beyond the GZK bound observed over the last few decades. We note that 
the model in its present form predicts that all of the super-GZK events are 
relatively well clustered in energy 
just beyond $\sim 10^{20}$ eV and that essentially 
no events should exist beyond those 
induced near the $Z$ pole. Obviously, if such `ultra'-GZK events were observed 
then there must be new processes which can also lead to enhanced annihilation 
cross sections beyond those arising in the SM.

In the RS model with the SM gauge and matter fields 
lying on the TeV brane there exist a 
Kaluza-Klein tower of massive, weak scale gravitons, $G^{(n)}$, with 
essentially 
electroweak couplings. There are basically two parameters in this model: 
the ratio $c=k/\mpl$, with $k$ a mass parameter with a magnitude 
comparable to the five-dimensional Planck scale, 
and the mass of the lowest lying graviton state. The masses of the tower KK 
states relative to the first non-zero mode are given by the 
ratio of roots of the Bessel function $J_1$ and $c$ is expected to lie in the 
range $\sim 0.01-1${\cite {RS,dhrprl,dhrprd}}. Specifically, 
while the zero mode graviton couples with a 
strength $\mpl^{-1}$, all of the remaining KK tower states couple as 
$\Lambda_\pi^{-1}$ where $\Lambda_\pi=\mpl ~e^{-\pi kr_c}$. For values of 
$kr_c$ in the range 11-12, the RS model provides a solution to the hierarchy 
problem. The masses of the KK states, $G^{(n)}$, are then given by 
$m_n=kx_n ~e^{-\pi kr_c}$ with $x_n$ being the $n^{th}$ roots of $J_1$. This 
implies that the 
tower mass spectrum is completely determined once the mass of the lowest lying 
excitation is known and is given by $m_n=m_1 x_n/x_1$. Thus we see that the 
parameters $c=k/\mpl$ and $m_1$ determine all of the other quantities within 
the RS model. 

Both phenomenological and theoretical constraints can 
be used to restrict this two dimensional model parameter space as has been 
discussed in our previous works{\cite {dhrprl,dhrprd}}. As an initial 
numerical example of the Gravi-burst mechanism 
let us consider the specific 
case where $m_1=600$ GeV and $k/\mpl=0.1$ which is a point in parameter 
space that is allowed by all of the 
current constraints; it is then straightforward to consider the more general 
case. For these specific values of the parameters we can calculate the cross 
section 
for $\nu \bar \nu$ annihilation into hadrons. This cross section now has 
a number of 
distinct contributions besides those arising from $q\bar q$ final states as 
in the decay of the $Z$ in the SM. (Even in the SM, above threshold, neutrinos 
can annihilate into pairs of $W$'s, $Z$'s and top quarks which can subsequently 
hydronically decay.) In the RS case, the gravitons not only lead to $q\bar q$, 
$t\bar t$, $W^+W^-$ and $ZZ$ final states but also to pairs of gluons and 
Higgs bosons, $gg,hh$. Gluons fragment directly into hadrons while the SM 
Higgs bosons decay mostly to $b\bar b$. 
(Here we assume for numerical purposes that 
the mass of the Higgs is 120 GeV.) By combining all of these individual 
process cross sections, including interference with SM $Z$ exchanges, 
we can calculate the full energy dependence of the total $\nu \bar \nu \to$ 
hadrons cross section. This allows us to determine the {\it ratio} of expected 
cosmic ray rates for super- and ultra-GZK events in units of the $Z$-pole 
induced rate $F_Z$ computed above. It is important to note that in forming 
this ratio almost all of the astrophysical uncertainties cancel except for 
the energy dependence of the neutrino flux. We find
\begin{equation}
R(\sqrt s)={F_{SM+GRAV}(\sqrt s)\over {F_Z}}={{2{\sqrt s}~\sigma_{ann}^{SM+GRAV}
(\sqrt s)~
(M_Z/\sqrt s)^\lambda}\over {M_Z^2 <\sigma_{ann}>^Z}} \,,
\end{equation}
where we have assumed that the neutrino spectrum above the the resonant $Z$ 
pole energy falls in a power-like manner as $\sim E_\nu^{-\lambda/2}$. 
We denote 
by $F_{SM}$ the complete energy dependent flux anticipated in the SM beyond 
that obtained through the use of the narrow-width approximation alone. 
(In what 
follows it will be sufficient to assume that this power-like fall off 
adequately describes the neutrino spectrum for a few orders of magnitude in 
energy above $E^{R_Z}_\nu$.) Integration 
of $R$ over a range of $\sqrt s$ values then tells us the relative rate of 
events expected in the RS model to those originating from $Z$-bursts.

\nn
\begin{figure}[htbp]
\centerline{
\psfig{figure=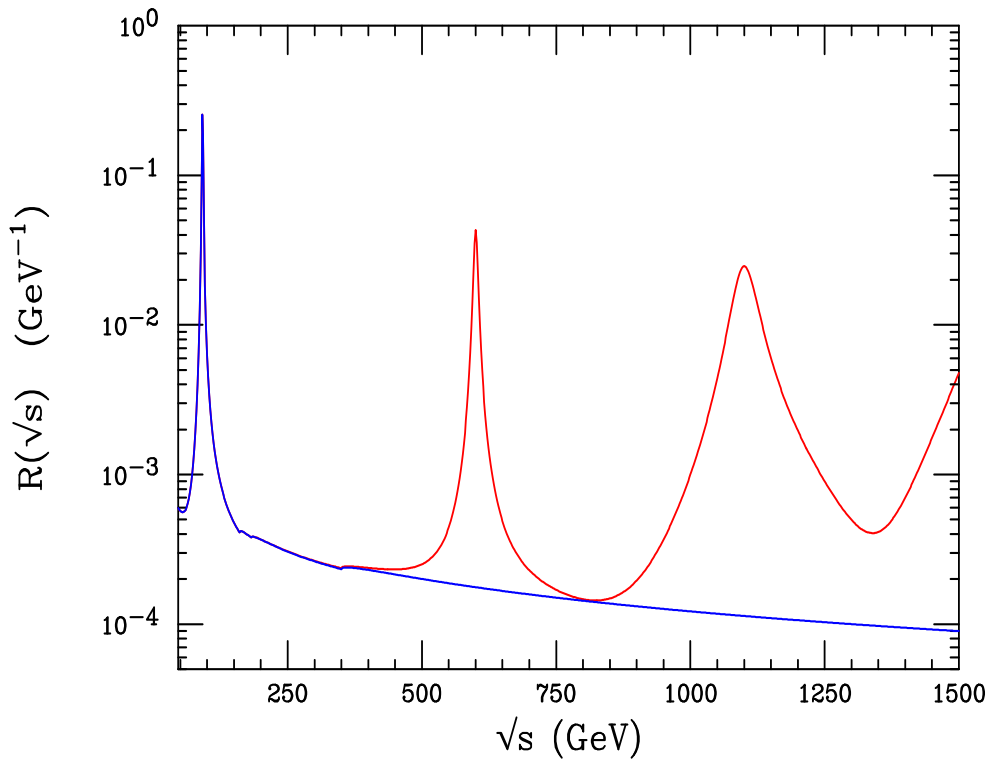,height=10cm,width=13cm,angle=90}}
\caption{Energy weighted total cross section for hadron production in units 
of that for the $Z$ pole in the Weiler $Z$-burst model for $\lambda=0$ as a 
function of center of mass energy for the SM(the relatively flat lower blue 
curve) and in the RS model(the 
upper red curve) with 
$c=k/\mpl=0.1$ and $m_1=600$ GeV. The small irregularities in the curves are 
due to $WW$, $ZZ$, $hh$ and $t\bar t$ thresholds.}
\label{fig1}
\end{figure}

To get an idea of what this ratio looks like as a function of energy we show 
the simplest specific case where $\lambda=0$ in 
Fig.~1. Note that the integral of $R$ under the $Z$ pole  gives the value unity 
as it should to reproduce the Weiler results. We are also reminded by the 
figure that even in the SM there exists a long 
high-energy tail to this ratio. In the RS scenario, the $Z$ peak is followed 
by a number of ever widening graviton peaks which also yield reasonably large 
cross sections.  From the 
figure, one can see that if the neutrino flux falls off slowly 
enough with energy we should expect events at even higher energies than those 
observed at present assuming that the $Z$ accounts for the `usual' super-GZK 
events. We will make this assumption in what follows, \ie, that the $Z$-burst 
scenario explains the observed super-GZK events.  
Given that hadronic multiplicities grow only very slowly with $\sqrt s$, as 
is observed in 
$e^+e^-$ annihilation data, we would interpret events induced by 
Gravi-bursting on the first graviton resonance to result in hadronic 
effective primaries that have energies approaching $10^{22}$ eV. As of yet no 
such events have been recorded which places bounds on 
the allowed parameters of the RS model for different values of the neutrino 
energy spectrum described by the parameter $\lambda$.

\section{Analysis}

As we found in the last section, under the assumption that $Z$-bursts explain 
the super-GZK events the existence of even higher energy ultra-GZK events is 
a rather generic prediction of the RS model. Let us restrict ourselves to the 
region $\sqrt s \geq 300$ GeV which corresponds to effective primary 
energies in excess of $10^{21}$ eV of which none have been yet observed. 
Integrating $R$ above this lower bound, even in the SM, can yield some 
`background' events; the use of the narrow width approximation 
is not strictly correct in that some rare events can arise from values of 
$\sqrt s$ away 
from the $Z$ pole. In the RS case, we integrate $R$ over the 
region from 300 GeV up to $\sqrt s=4m_1$ beyond which perturbation theory 
fails. {\footnote {We note that the RS model as described in four dimensions 
is a non-renormalizable theory. In addition, once the value of $\sqrt s$ 
significantly exceeds 
$\Lambda_\pi$ the theory also becomes non-perturbative and only qualitative 
statements can be made about the behavior of the cross 
section{\cite {dhrprl}}.} 
This yields a conservative lower bound on the total number of ultra-GZK 
events that are predicted in the RS model since most certainly more events can 
arise from even larger values of $\sqrt s$. Integrating 
$R$ in the SM over the above ranges and assuming that the 20 super-GZK events 
are from the $Z$-pole region, we find that for $\lambda=1(2,3)$ we would 
expect to have already seen $\simeq 0.24(0.04,0.008)$ ultra-GZK background 
events from the tail of the $Z$ pole, which is quite acceptable. (As discussed 
above we might expect that $\lambda \geq 1.8$ is allowed by Fly's Eye data 
if the energy dependence of the 
neutrino spectrum below and above the $Z$-pole are similar.) Performing 
the same calculation in the RS model for a fixed set of values of $m_1$ and 
$\lambda$ it will be clear that for some range of $k/\mpl$ the cross sections 
will be too large to have avoided the present  
non-observation of ultra-GZK events. In 
the usual manner this means that we can place a $95\%$ CL bound on $k/\mpl$ 
as a function of $m_1$ for different assumed values of $\lambda$, using the 
existing data.

\nn
\begin{figure}[htbp]
\centerline{
\psfig{figure=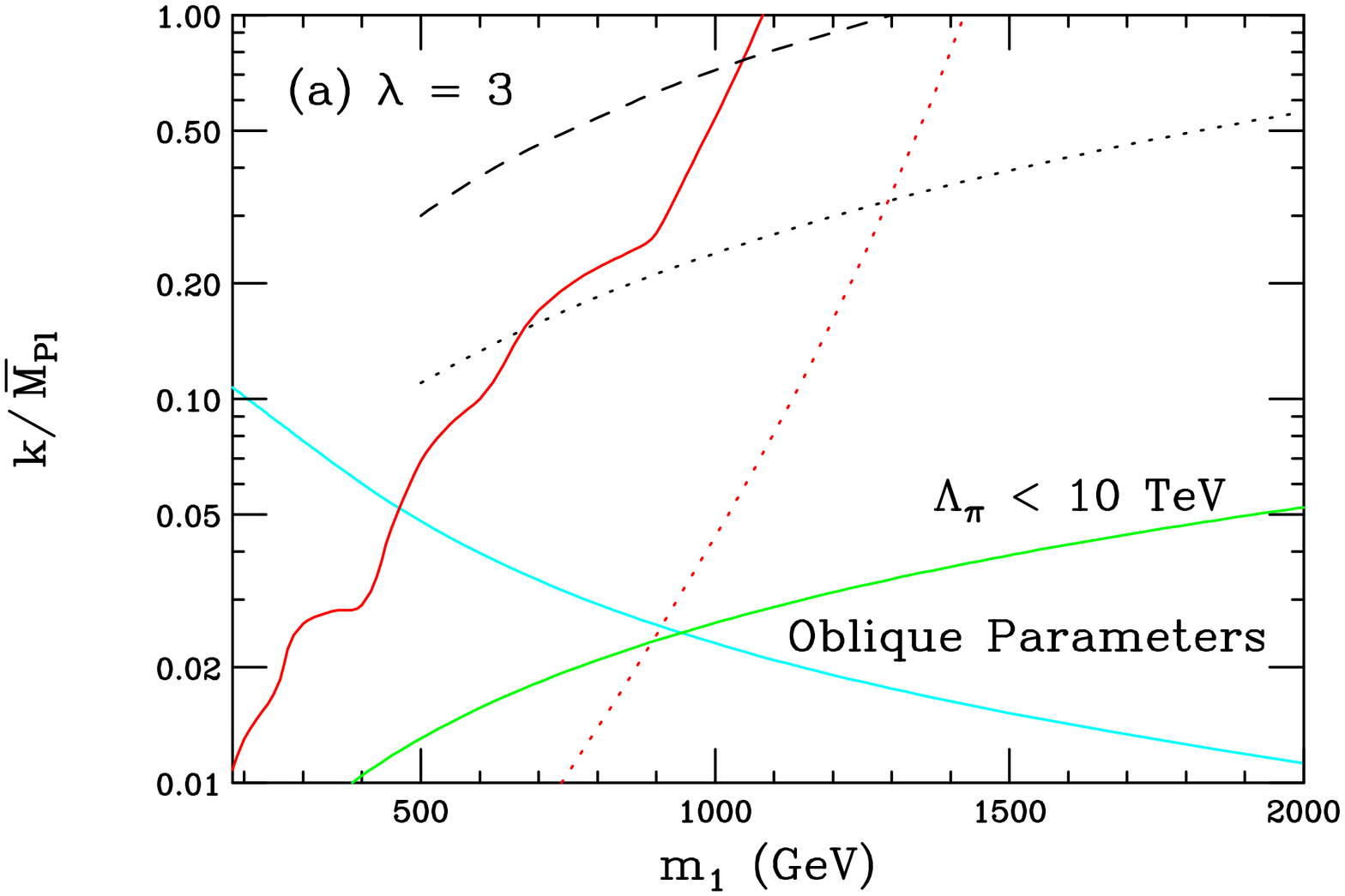,height=8.cm,width=8cm,angle=90}}
\vspace*{0.25cm}
\centerline{
\psfig{figure=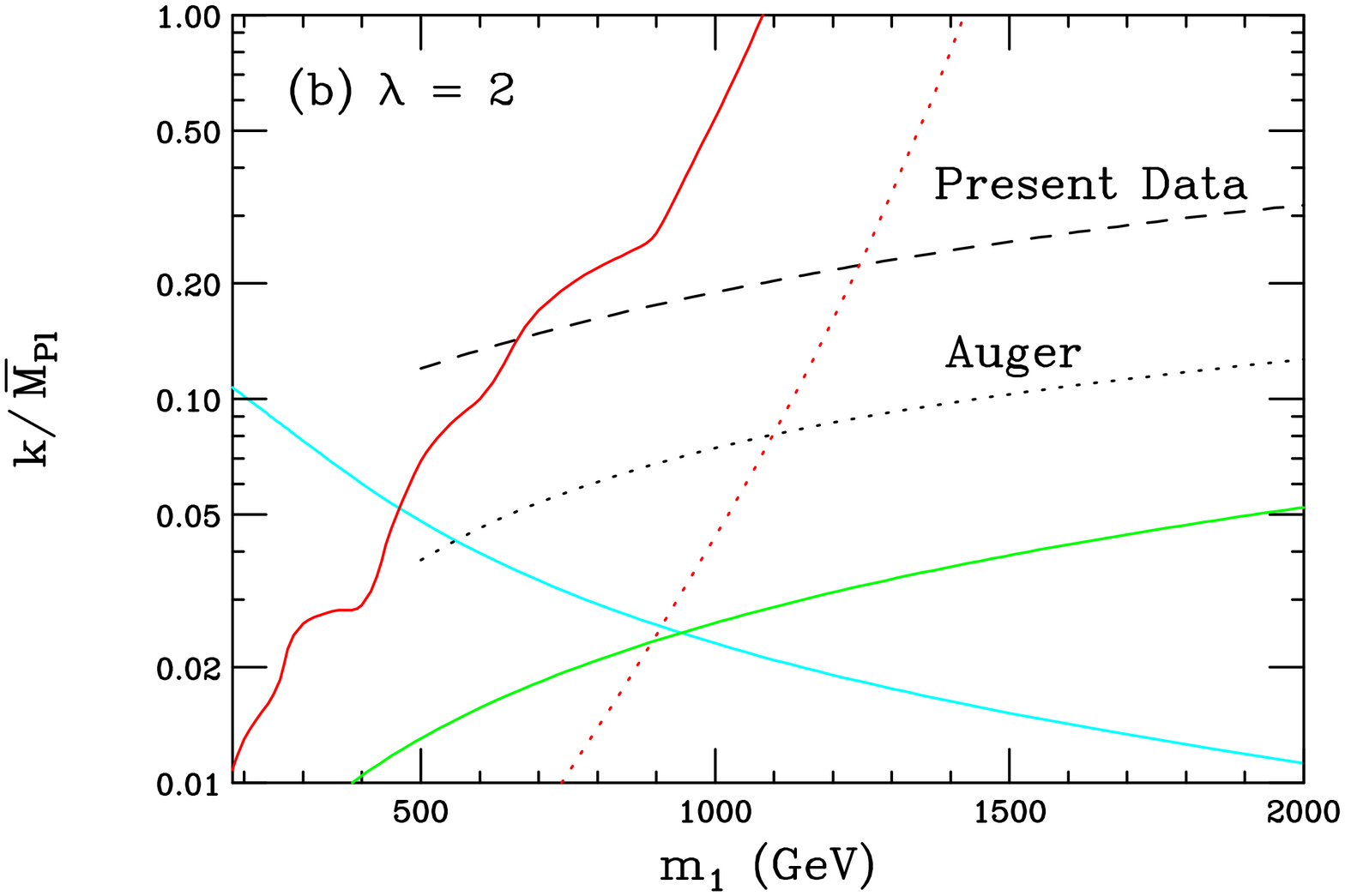,height=8.cm,width=8cm,angle=90}
\psfig{figure=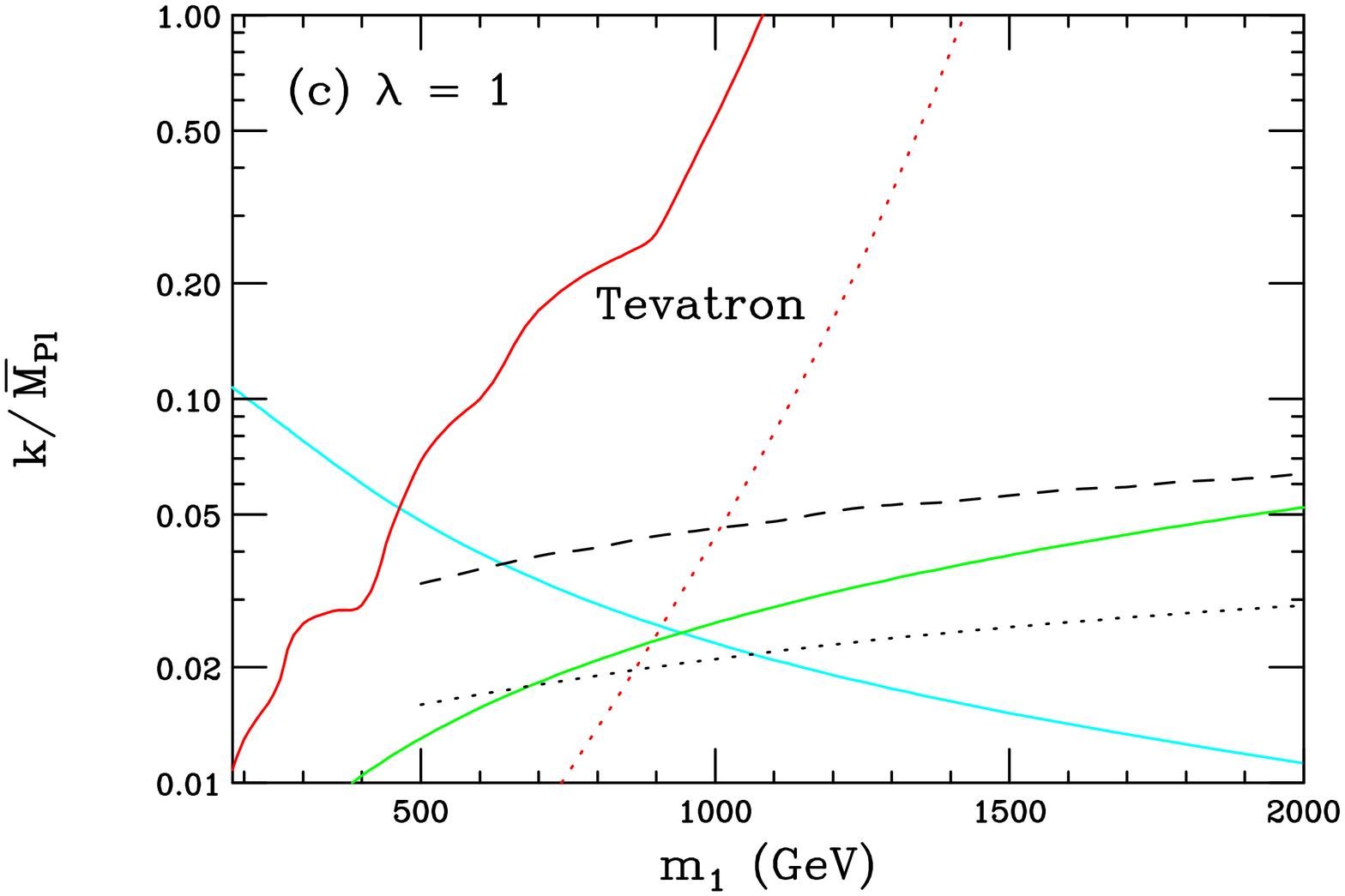,height=8.cm,width=8cm,angle=90}}
\vspace*{0.25cm}
\caption{Allowed region in the ($k/\mpl$)-$m_1$ plane. The solid(dotted) 
diagonal red 
curve excludes the region above and to the left from direct searches for 
graviton resonances at the 
Run I(II, $30~fb^{-1}$) Tevatron. The light blue(green) curve is an indirect 
bound from the oblique 
parameter analysis (based on the hierarchy requirement that $\Lambda_\pi<10$ 
TeV) and excludes the region below it. The 
black dashed(dotted) curves excluding the regions above them 
at $95\%$ CL based on present (anticipated future Auger) cosmic ray data. The 
top(bottom left, bottom 
right) panel corresponds to $\lambda=3(2,1)$ which describes the fall with 
energy of the neutrino flux as $E^{-\lambda/2}$.}
\label{fig2}
\end{figure}

The dashed curves in Fig.~2 show the results of this analysis using the 
existing data for various values of $\lambda$; other constraints obtained on 
the ($k/\mpl$)-$m_1$ parameter space from our earlier work{\cite {dhrprd}} are 
also shown. We 
see immediately that the effectiveness of the bound is quite strongly dependent 
on the value of $\lambda$. For $\lambda \geq 3$ at most only a tiny region 
beyond that excluded by existing Tevatron Run I data is now ruled out. As 
$\lambda$ decreases the size of the region presently excluded by cosmic ray 
data grows rapidly. For $\lambda=2$ a substantial region allowed by the 
present Tevatron data becomes excluded. Furthermore, a sizeable region 
{\it beyond} that accessible at Run II with a luminosity of 30 $fb^{-1}$ is 
also excluded. Using accelerators alone this region would be inaccessible 
until after the LHC turns on but here we see that it would be excluded by 
cosmic ray data provided $\lambda \leq 2$. For $\lambda \leq 1$ the bound is 
extremely powerful and at most only a tiny sliver of the RS model parameter 
space would remain viable. 

In the region below the dashed curves, which is not excluded by 
existing cosmic ray 
data, we might expect ultra-GZK events to show up in future experiments at 
reasonable rates. If this does not happen it's clear that the present bounds 
discussed above will improve drastically especially with the new cosmic ray 
observatories such as Auger{\cite {Auger}} coming on line. Within 
a 5 year period of data taking at Auger 
one would expect $\sim 1000$ super-GZK events{\cite {Cronin}} induced by 
$Z$-Bursts with correspondingly higher sensitivity to the ultra-GZK region. 
If no events above the SM background from the $Z$ pole tail are observed at 
Auger during this period we can repeat the analysis above to obtain strong 
constraints on the RS parameter space as shown by the dotted curves in Fig.~2. 
We find the SM background expectations in this case for $\lambda=1(2,3)$ to 
be $\simeq 12.5(2.05,0.422)$ events. 
Here we see that for $\lambda=2,3$ the size of the presently allowed region 
is quite significantly reduced. Particularly note the case 
$\lambda \leq 1$ where we find that the non-observation of any ultra-GZK 
events at Auger would completely {\it exclude} the RS model with the SM gauge 
and matter fields on the wall. This is a very powerful result. 

If events above background are observed at Auger due to gravi-bursts they 
will have two distinctive characteristics. First, due to the resonance 
structure predicted by the RS model the energies of the effective hadronic 
primaries will show peaking at a set of fixed energies provided the energy 
resolution of the detectors is sufficiently good. Second, in addition to the 
rather `soft' photons arising from conventional fragmentation $\pi^0$'s, 
much harder photons can arise from the direct decays of the gravitons in the 
KK tower. As shown in our earlier work, gravitons in the mass range of interest 
can decay with a reasonable branching fraction, $\simeq 4-5\%$, into photons. 
Since they carry half of the energy of the resonance mass these photons will 
have energies an order of magnitude or more larger than those 
arising from $\pi^0$ decays. Of course with this rather small branching 
fraction a reasonable number of more `ordinary' ultra-GZK events should be 
observed before one induced by these very hard photons. 

So far we have only discussed the case of a single massive neutrino; data 
based on oscillation solutions to the solar neutrino problem{\cite {SuperK}} 
suggest a second massive state exists near $3\times 10^{-3}$ eV, about a 
factor 20 or so in mass below the 0.06 eV case discussed above. This second 
neutrino can also induce a $Z$-burst but only if the energy of the 
corresponding incident neutrino is $\simeq 20$ times larger, 
$\simeq 1.4\times 10^{24}$ eV. In comparison to that 
for the case of the higher mass neutrino the flux of these lighter 
neutrinos would be $\simeq (20)^{\lambda/2}$ times smaller. This assumed 
neutrino mass ratio, strongly suggests that $\lambda \geq 2$ to avoid 
ultra-GZK events from the second $Z$ resonance. To eliminate any additional 
background generated by this new $Z$ contribution, we would need to raise 
lower bound on our $\sqrt s$ integration, which is actually an integral over 
the neutrino energy. (The $\sqrt s=300$ GeV lower bound translates into a 
minimum neutrino energy of $\simeq 8\times 10^{23}$ eV and thus would now 
include additional background events from the second $Z$ peak.) Raising the 
minimum neutrino energy by a factor of two would remove the second $Z$ 
contribution while still staying comfortably below the excitation energies of 
any of the gravitons. In this case we would expect at most only slight 
alterations in the bounds presented above.

Before we conclude we briefly discuss a generalization of the RS model where 
the SM gauge and matter fields are taken off the TeV brane{\cite {dhrprd}} 
and how this would 
influence our results above. In this case not only can graviton towers be 
exchanged in the $\nu \bar \nu \to$ hadrons process but now there are also 
$Z$ boson towers whose members are generally interspaced in mass with the 
gravitons. If these additional contributions are also present it is quite 
possible that the neutrino cross section can be significantly enhanced leading 
to even stronger limits than those obtained above using current data. One 
might also expected that with increased cross sections it might be possible 
to probe cases where the slope of the neutrino energy spectrum is even steeper 
than what we have considered here. 
Unfortunately, to determine how much our previous results are modified in 
a quantitative manner requires a detailed analysis which is far 
beyond the scope of this paper.

\section{Conclusions}

In this paper we have examined the possible contribution to the spectrum of 
cosmic rays beyond the GZK cutoff due to new physics arising in the 
Randall-Sundrum model of localized gravity. 
Our analysis is based on the assumptions ($i$) that 
the events observed immediately above the GZK bound can be explained by the 
$Z$-Burst mechanism and ($ii$) the neutrino spectrum needed for $Z$-Bursts 
extends s few orders of magnitude further in neutrino energy with a reasonably 
slow fall-off. If these conditions hold then the existence of a series of 
$s$-channel Kaluza-Klein graviton resonances in the $\nu \bar \nu \to$ hadrons
channel, which is 
predicted in the RS model, can lead to events with even higher 
energies, ultra-GZK, due to Gravi-Bursts. The rate for these bursts are 
generally at or near the present level of observability for a wide range of 
RS model parameters. The fact that such events are not as yet observed can be 
used to constrain the parameter space of the RS model once a specific form of 
the neutrino energy spectrum is assumed. These bounds can be more restrictive 
than those that can be obtained from the lack of graviton resonance 
production at the Tevatron during Run II ($30~fb^{-1}$) if 
the fall-off with energy of the UHECR neutrino flux is linear or less steep. If 
ultra-GZK events are not observed by future experiments such as the Auger 
Array, then the resulting bounds on the RS model can be complementary to those 
obtainable at the LHC. If such events are observed at future experiments, the 
RS resonance structure may be observable given both sufficient statistics and 
good hadronic energy resolution. In addition to hadronic modes, the RS graviton 
KK tower states can directly decay to photon pairs which will have more than 
an order of magnitude greater energies that those that can arise due to 
ordinary fragmentation into $\pi^0$'s which subsequently decay into two 
photons. If photon and hadron induced showers can be distinguished at 
such energies this will provide a unique signature for the 
RS model as the origin of the ultra-GZK events.

\noindent{\Large\bf Acknowledgements}

The authors would like to thank the Aspen Center for Physics, where this work 
was begun, for its hospitality. T.G.R. would like to thank T. Weiler and 
J.L.H. would like to thank E. Kolb for discussions related to this paper.

\newpage

%
\def\MPL #1 #2 #3 {Mod. Phys. Lett. {\bf#1},\ #2 (#3)}
\def\NPB #1 #2 #3 {Nucl. Phys. {\bf#1},\ #2 (#3)}
\def\PLB #1 #2 #3 {Phys. Lett. {\bf#1},\ #2 (#3)}
\def\PR #1 #2 #3 {Phys. Rep. {\bf#1},\ #2 (#3)}
\def\PRD #1 #2 #3 {Phys. Rev. {\bf#1},\ #2 (#3)}
\def\PRL #1 #2 #3 {Phys. Rev. Lett. {\bf#1},\ #2 (#3)}
\def\RMP #1 #2 #3 {Rev. Mod. Phys. {\bf#1},\ #2 (#3)}
\def\NIM #1 #2 #3 {Nuc. Inst. Meth. {\bf#1},\ #2 (#3)}
\def\ZPC #1 #2 #3 {Z. Phys. {\bf#1},\ #2 (#3)}
\def\EJPC #1 #2 #3 {E. Phys. J. {\bf#1},\ #2 (#3)}
\def\IJMP #1 #2 #3 {Int. J. Mod. Phys. {\bf#1},\ #2 (#3)}

\end{document}